\documentclass[twocolumn,astrosymb,times,tighten,resetfootnote,twocolappendix]{aastex701} 
\usepackage{amsmath}
\usepackage{placeins} 

\begin{document}
\title{Filament-Arm Node Systems (FANS):\\A new probe of the winding and chirality of the cosmic web}

\author[orcid=0000-0003-4936-0069,sname='da Silveira Ferreira']{Pedro da Silveira Ferreira}
\affiliation{Center for Cosmology and Computational Astrophysics, Institute for Advanced Study in Physics, Zhejiang University, Hangzhou 310058, China.}
\affiliation{Institute of Astronomy, School of Physics, Zhejiang University, Hangzhou 310058, China}
\email[show]{dasferreira.pedro@gmail.com}  

\author[orcid=0000-0001-8531-9536,sname='Cen']{Renyue Cen} 
\affiliation{Center for Cosmology and Computational Astrophysics, Institute for Advanced Study in Physics, Zhejiang University, Hangzhou 310058, China.}
\affiliation{Institute of Astronomy, School of Physics, Zhejiang University, Hangzhou 310058, China}
\email[show]{renyuecen@zju.edu.cn}  

\begin{abstract}
We define Filament-Arm Node Systems (FANS), a new class of cosmic-web objects consisting of multiple filament arms connected to a common node, and use their geometry to construct an object-based test of cosmic chirality. With the cosmic-web node providing a natural reference point, FANS span characteristic scales of $\sim 60\,h^{-1}{\rm Mpc}$, allowing the coherent winding of their arms around the node to be measured directly. We define cosmic spirality as the amplitude of this winding and handedness as its signed clockwise (CW) or counter-clockwise (CCW) sense. We further introduce an axis-conditioned extrinsic estimator that searches for the projected signature of a coherent three-dimensional axial pattern of the handedness field, corresponding to a preferred axis of winding. Applying these measurements to two distinct SDSS filament reconstructions, and assessing their significance with sign-flip nulls and 27 controlled variations of the FANS construction parameters, we find no significant preference for CW or CCW winding, either globally or around a preferred axis. The data are therefore consistent with parity symmetry, as expected in standard $\Lambda$CDM cosmology. Comparing the lowest- and highest-redshift bins over $0.1<z<0.6$, we find that cosmic spirality decreases with cosmic time, a trend that also persists within individual arm-multiplicity classes and may reflect the gravitational straightening of filament arms. These first constraints establish FANS as a new probe of cosmic-web symmetry and large-scale morphology and provide a baseline for future parity tests with the much larger filament samples expected from upcoming surveys.
\end{abstract}

\keywords{\uat{Cosmic web}{330}; \uat{Large-scale structure of the universe}{902}; \uat{Cosmological evolution}{336}}

\section{Introduction}
\label{sec:intro}

Symmetries are among the most fundamental assumptions in cosmology. The standard cosmological model is built on the assumption that, on sufficiently large scales, the Universe is statistically homogeneous and isotropic \citep[e.g.][]{Clarkson:2010uz,Aluri:2022hzs}. A further, usually implicit, assumption is parity symmetry: the statistical distribution of cosmic structures is unchanged under spatial reflection, so that left- and right-handed configurations occur with equal probability \citep{Samandar:2024topology}. A detection of a preferred cosmic handedness would therefore be profound, pointing either to new parity-violating physics, non-trivial cosmic topology, or an observational systematic. Conversely, a null result provides a useful consistency test of the standard cosmological picture.

In standard $\Lambda$CDM, scalar density perturbations and their late-time gravitational evolution are not expected to generate an intrinsic parity-odd\footnote{Parity-even quantities are unchanged by spatial reflection, whereas parity-odd quantities change sign.} density field. The reason is that ordinary gravitational evolution does not distinguish a configuration from its mirror image. If the initial density field has no preferred handedness, the equations of motion evolve the original and reflected configurations in equivalent ways. At the primordial level, parity-odd scalar correlations require a source of chirality in the dynamics or initial state. In the standard generation of scalar perturbations, there is no such ingredient; as a result, parity-odd scalar correlations are strongly restricted \citep{Cabass:2022rhr}. However, parity violation is still possible through physical mechanisms including axion--gauge-field dynamics, dynamical Chern--Simons gravity, chiral gravitational waves, massive spinning fields during inflation, non-Bunch--Davies initial states, or helical primordial magnetic fields \citep[e.g.][]{Ozsoy:2021onx,Creque-Sarbinowski:2023wmb,Fujita:2023inz,Reinhard:2024evr,Yura:2025mus}. At the same time, observed galaxy clustering is not the primordial density field itself: galaxies are observed on the past light cone, in redshift space, through a finite survey mask and with non-trivial selection functions. Doppler, lensing, redshift-space, projection, and reconstruction effects can therefore generate or modulate apparent odd-parity contributions even when the underlying density field is parity symmetric \citep{Paul:2024uim}. This makes independent observables essential for testing cosmic parity.

Parity symmetry has been tested extensively with the cosmic microwave background (CMB). In a statistically isotropic and parity-invariant cosmology, the ensemble-averaged CMB $TB$ and $EB$ power spectra vanish. Gravitational lensing converts $E$ modes into $B$ modes and induces off-diagonal mode coupling, but preserves parity and therefore does not generate nonzero ensemble-averaged diagonal $TB$ or $EB$ spectra \citep{Hu:2000ee,Lewis:2006fu}. By contrast, cosmic birefringence, chiral primordial gravitational waves, and frequency-dependent Faraday rotation from a coherent primordial magnetic field can generate such signals \citep{LueWangKamionkowski1999,GluscevicKamionkowski2010,Kosowsky:1996yc,Kristiansen:2008tx}. Planck and ACT analyses have reported sub-degree rotation-angle estimates at $2$--$3\sigma$ significance, although their cosmological interpretation remains limited by polarization-angle calibration, Galactic dust $EB$, and other instrumental systematics \citep{MinamiKomatsu2020,DiegoPalazuelosEtAl2022,Diego-Palazuelos:2025dmh}. Meanwhile, parity-odd CMB trispectrum searches remain consistent with zero \citep{Philcox2023CMBParity,PhilcoxShiraishi2024CMBParity}. Thus, no established cosmological detection of CMB parity violation has yet emerged.

At lower redshift, one class of parity tests uses the apparent winding sense of spiral galaxies. Direct searches for a large-scale dipole in spiral-galaxy handedness have yielded mixed results. A Galaxy Zoo analysis found no significant dipole after accounting for human-classification bias and an overall handedness offset \citep{Land:2008vh}, whereas later SDSS analyses using randomly mirrored manual classifications or automated image processing reported dipolar asymmetries and preferred axes \citep{2011PhLB..699..224L,2012PhLB..715...25S}. Separately, Galaxy Zoo data yielded a tentative positive two-point correlation of spin chirality on sub-Mpc scales, probing local spin coherence rather than a global handedness dipole \citep{Slosar:2008xh}. Although galaxy spins are intuitive tracers of chirality and are closely connected to tidal-torque physics and the cosmic web \citep[e.g.][]{Porciani:2001db,Tempel:2013gqa,Dubois:2014lxa}, they are difficult to interpret as clean parity probes because spin reconstruction is sensitive to morphology, inclination, dust, imaging systematics, and projection effects, while baryonic evolution and environmental torques can modify any primordial signal.

Another approach uses parity-odd modes of the galaxy four-point correlation function (4PCF). Because a tetrahedron is the lowest-order three-dimensional configuration that is generically distinct from its mirror image, the 4PCF provides the lowest-order scalar density statistic sensitive to parity violation \citep{Philcox:2022hkh,Hou:2022wfj,Jamieson2024POP}. While initial BOSS analyses reported evidence for parity-odd modes, subsequent BOSS and DESI analyses, including a recent compressed-trispectrum test, found that the inferred significance depends sensitively on covariance modeling, data--mock consistency, and analysis choices, leaving the current observational picture inconclusive \citep{Krolewski:2024paz,Slepian:2025kbb,Hou:2025cey,Gao:2026zaz}.

This motivates parity-sensitive observables that compress different aspects of the same density field. The cosmic web offers such a possibility: by conditioning on nodes where multiple filaments meet, one can probe the winding geometry of connected filament arms rather than high-order correlations of the density field as a whole. Such an observable probes geometry on larger scales than galaxy spins, is less directly tied to baryonic angular-momentum physics, and provides a complementary parity test with different systematics. In addition, the recently reported Large-Scale Axial Intrinsic Alignment (LAIA), which provides evidence for a coherent preferred axis in galaxy and filament orientations \citep{daSilveiraFerreira:2025bep}, further motivates asking whether parity-sensitive cosmic-web observables may exhibit a similar preferred-axis organization, thereby testing whether parity-sensitive structure and preferred-axis statistical anisotropy may share a common origin.

In this Letter we introduce Filament-Arm Node Systems (FANS), configurations of multiple filament arms connected to a common cosmic-web node, as a new object-based probe of cosmic-web chirality. We define cosmic spirality, the parity-even amplitude of the projected azimuthal winding of the arms, and handedness, the parity-odd signed winding. We further develop an extrinsic, axis-conditioned estimator that searches for the direction about which handedness becomes maximally coherent after accounting for the sign reversal expected from a three-dimensional axial winding pattern. This extends the analysis beyond a sky-averaged parity test to an axis-dependent test of statistical isotropy. We apply these estimators to two distinct SDSS filament catalogues, assess their significance using sign-flip nulls, and test robustness under 27 variations of the FANS construction parameters. Our analysis provides the first object-based constraint on cosmic-web handedness and an independent test of parity-sensitive structure reported in galaxy-spin studies and higher-order density statistics.

\section{Filament-Arm Node Systems (FANS)}
\label{sec:fans}
\vspace{0.2cm}

Filament-Arm Node Systems (FANS) are systems of two or more filament arms that emanate from a common cosmic-web node and remain coherently connected to it over a finite radial range. The node provides a physically motivated origin and defines an outward radial orientation for each connected arm. This removes the sign ambiguity of otherwise unoriented filament ridges, allowing clockwise (CW) and counter-clockwise (CCW) azimuthal winding to be distinguished. This makes FANS natural systems for measuring the amplitude and handedness of cosmic-web winding.

Cosmic-web connectivity differs from local filament multiplicity because the number of branches can change with distance from a node through bifurcations \citep{Codis2018Connectivity}. Connectivity also correlates with node and galaxy properties \citep{DarraghFord2019Connectivity,Kraljic2020Connectivity}. FANS multiplicity is the number of arms retained by the tracing procedure. If typical arm geometries differ between multiplicities, changes in their relative abundance could contribute to a spirality--redshift trend. We therefore analyse the multiplicity classes separately to distinguish geometric evolution within these classes from changes in their population mixture.

We apply this construction to two SDSS-based filament reconstructions. The first is the SDSS DR12 filament catalogue of \citet{Chen:2015oqa} (Chen), which provides both filament ridge points and cosmic-web nodes in redshift slices, enabling a direct FANS construction. We then use the SDSS DR16 filament catalogue of \citet{Duque:2021xgw} (Carrón), which identifies filament ridges using a machine-learning-assisted implementation of the Subspace-Constrained Mean-Shift (SCMS) method \citep{Chen:2015ofa}. Because the public Carrón catalogue provides filament ridges but not the corresponding nodes, we use a SCONCE reconstruction of the same region and redshift shells \citep{Zhang:2022yfi} solely to define the node positions. We adopt the SCONCE parameter choices used in the validation tests of \citet{Zhang:2022yfi}. The Chen and Carrón reconstructions use different SDSS data releases, footprints, and implementations of density-ridge filament finding, providing a useful cross-check of reconstruction-dependent systematics. The Chen sample is restricted to the northern SDSS footprint, whereas the Carrón sample includes both northern and southern SDSS regions.

\begin{figure}
\centering
\hspace{-0.8cm}\includegraphics[width=0.68\linewidth]{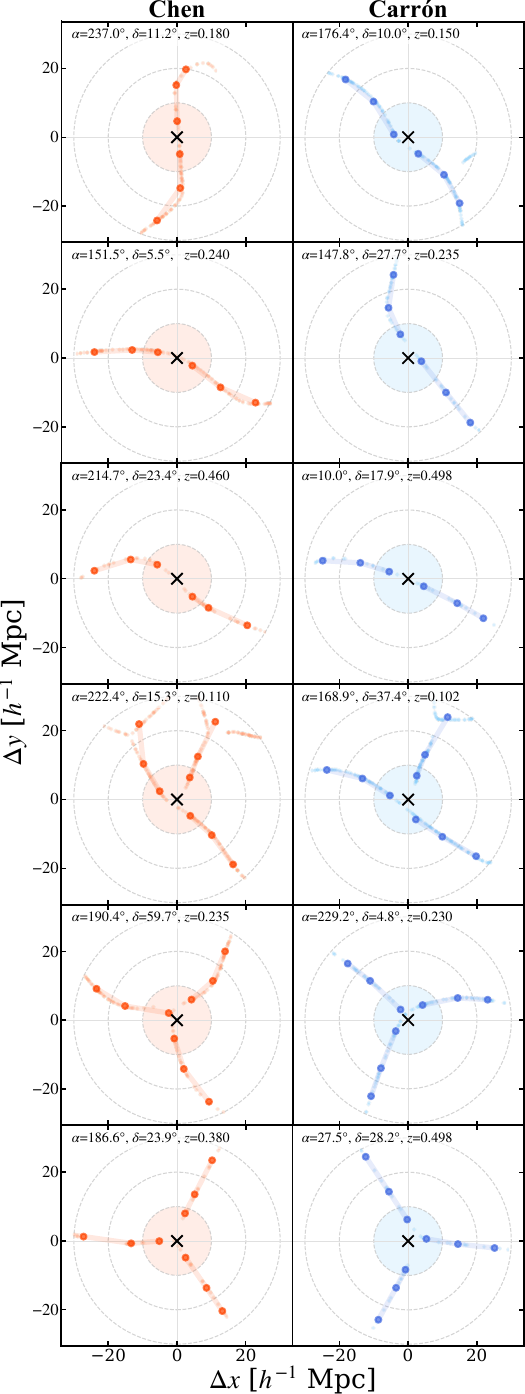}
\caption{Representative two- and three-arm FANS from the Chen and Carrón samples. Small dots show the filament-ridge points; $\times$ symbols mark the cosmic-web nodes. The lines connecting large circles trace the reconstructed arms obtained using the fiducial construction parameters. The shaded circle marks the linking region, within which filament-ridge points are used to seed candidate arms.}\label{fig:fans_examples}
\end{figure}

Operationally, for each node we search for neighbouring ridge points within a projected comoving radius $R_{\rm arm}=30\,h^{-1}{\rm Mpc}$. Here a ridge point denotes a discrete sampling point of the reconstructed filament ridge. Candidate arms are seeded by ridge points within $R_{\rm link}=10\,h^{-1}{\rm Mpc}$ of the node and grouped according to their node-centred polar angles. The arm half-opening angle $\Delta\phi_{\rm arm}$ is the maximum angular offset allowed between a ridge point and the direction of a candidate arm: points within $\pm\Delta\phi_{\rm arm}$ can be assigned to the same arm. Our fiducial value is $\Delta\phi_{\rm arm}=35^\circ$; this is an association tolerance rather than a physical filament width. Each arm is then traced radially outward through $N_{\rm seg}=3$ segments of length $L_{\rm seg}=10\,h^{-1}{\rm Mpc}$. Representative two- and three-arm FANS from both filament catalogues are shown in Fig.~\ref{fig:fans_examples}, illustrating the nodes, filament ridges, radial segmentation, and resulting FANS geometry.

Within this scale, visual inspection indicates that arms connected to a node are typically still morphologically stable, with fragmentation, bifurcation, merging, or termination becoming more frequent at larger radii. Each radial segment must contain at least three ridge points. Tracing stops at the first failed segment; accepted arms must contain at least two consecutive valid segments, and each FANS must contain at least two valid arms. The latter requirement ensures that the observable characterizes a multi-arm node environment rather than the geometry of a single isolated ridge. Segment positions and local tangent directions are computed using the local reconstructed density assigned to each ridge point as a weight, with the weights normalized within each segment, so that better-supported regions of the ridge contribute more strongly to the measured arm geometry.

The fiducial construction parameters were selected from visual inspection of 1000 representative systems, without reference to the measured handedness or its statistical significance, and tested across a 27-point robustness grid varying $L_{\rm seg}\in\{7.5,10,12.5\}\,h^{-1}{\rm Mpc}$, $R_{\rm link}\in\{8,10,12\}\,h^{-1}{\rm Mpc}$, and $\Delta\phi_{\rm arm}\in\{30^\circ,35^\circ,40^\circ\}$ (Appendix~\ref{app:robustness}). These ranges test the main reconstruction trade-offs: narrower opening angles or shorter radial segments increase sensitivity to ridge fragmentation and small-scale sampling noise, whereas broader angles or longer segments can merge neighbouring arms or wash out the arm curvature of interest.

\begin{figure}[!ht]
\centering
\includegraphics[width=0.915\linewidth]{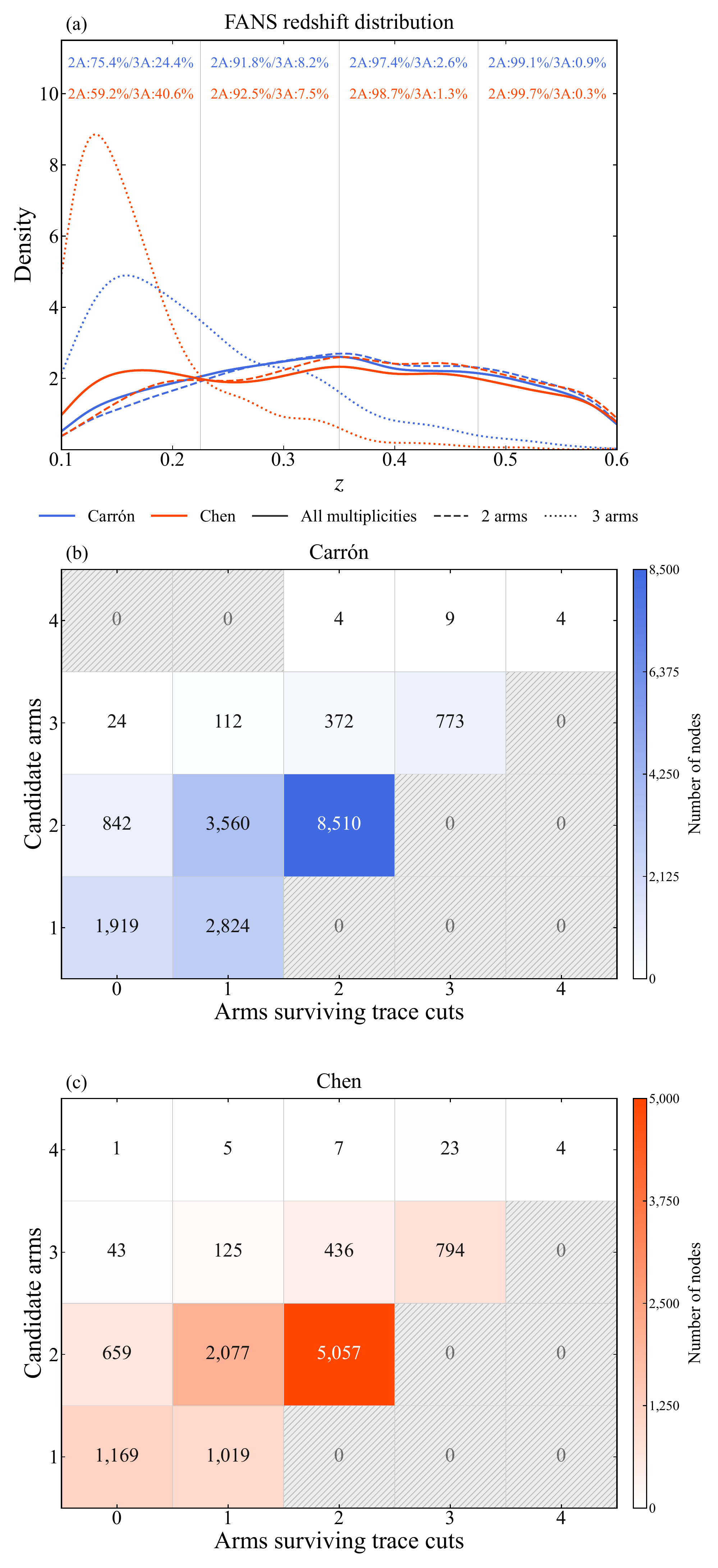}
\caption{FANS redshift distributions and arm multiplicity. (a) Separately normalized distributions for all accepted FANS and the two- and three-arm subsamples; annotations give their fractions among all accepted FANS in each redshift bin. (b,c) Changes in arm multiplicity through the tracing quality cuts for the Carr\'on and Chen samples, respectively.}\label{fig:fans_redshift}
\end{figure}

For the final FANS analysis, we restrict both catalogues to the common redshift range $0.1<z<0.6$. Although both catalogues contain individually well-reconstructed filament segments outside this interval, the number density of usable nodes and connected ridges becomes too low at $z<0.1$ and $z>0.6$ to yield  statistically stable FANS samples. We divide this interval into four redshift bins with edges $z=\{0.1,0.225,0.35,0.475,0.6\}$. The redshift distributions and multiplicity-selection diagnostics are shown in Fig.~\ref{fig:fans_redshift}. To reduce boundary-induced distortions in the reconstructed filaments, we retain only systems whose nodes lie at least $70\,h^{-1}{\rm Mpc}$ in projected comoving distance from the angular survey boundary at the node redshift. This conservative cut mitigates spurious arm incompleteness or curvature induced by the finite survey mask. After all quality cuts, the  Chen sample contains 6321 FANS: 5500 with two arms, 817 with three arms, and four with four arms, totalling 13467 arms. The Carrón sample contains 9672 FANS: 8886 with two arms, 782 with three arms, and four with four arms, totalling 20134 arms. Neither fiducial sample contains a system with five or more arms.

The three-arm fraction decreases from 24.4\% to 0.9\% between the first and last bins in the Carrón sample and from 40.6\% to 0.3\% in the Chen sample. Figure~\ref{fig:fans_redshift} also compares arm multiplicities before and after the tracing quality cuts. Among accepted two-arm FANS, 4.2\% in the Carrón sample and 8.1\% in the Chen sample originated from candidates with at least three arms. Thus tracing can reduce multiplicity, but the predominance of two-arm systems is not produced solely by such reductions.

\section{FANS Spirality and Handedness}
\label{sec:spirality}

FANS turn the geometry around a cosmic-web node into a natural local coordinate system: the node defines the origin, while its connected arms trace curves that can be decomposed into radial and azimuthal directions. This turns the visual notion of winding into two simple questions: how strongly do the arms depart coherently from the radial direction, and is that departure preferentially CW or CCW? Unlike galaxy-spin chirality, which is inferred from the apparent morphology of the spiral arms and is commonly compressed into a binary CW/CCW label in cosmological tests, FANS retain both the sign and continuous magnitude of coherent winding. Because their arms are traced over tens of megaparsecs and through multiple radial segments, this coherent signal can be distinguished from local ridge fluctuations or small-scale bends.

For each arm $i$ and valid radial segment $k$, we work in the local projected plane centred on the node. We define $\hat{\mathbf r}_{ik}=(\hat r_{x,ik},\hat r_{y,ik})$ as the unit radial direction from the node to the density-weighted segment position. Because a filament ridge has no intrinsic orientation, its tangent initially has an arbitrary sign. We remove this ambiguity by orienting the local tangent $\hat{\mathbf t}_{ik}$ away from the node, requiring $\hat{\mathbf t}_{ik}\cdot\hat{\mathbf r}_{ik}>0$. The corresponding counter-clockwise azimuthal direction and local signed winding, defined in the tangent plane (image plane), are then
\begin{equation}
\hat{\boldsymbol{\phi}}_{ik}
=
(-\hat r_{y,ik},\hat r_{x,ik}),
\qquad
s_{ik}
=
\hat{\mathbf t}_{ik}\cdot\hat{\boldsymbol{\phi}}_{ik}.
\label{eq:segment_spirality}
\end{equation}
A purely radial segment has $s_{ik}=0$, while increasingly tangential segments approach $|s_{ik}|=1$. The sign carries the projected winding sense: $s_{ik}>0$ corresponds to CCW and $s_{ik}<0$ to CW winding around the node. Writing $\theta_{r,ik}$ and $\theta_{t,ik}$ for the position angles of $\hat{\boldsymbol r}_{ik}$ and $\hat{\boldsymbol t}_{ik}$, Eq.~(1) may equivalently be written as $s_{ik}=\sin(\theta_{t,ik}-\theta_{r,ik})$. An explicitly axial alternative is the projected two-dimensional pseudoscalar $\chi_{ik}\equiv\sin\!\left[2\left(\theta_{t,ik}-\theta_{r,ik}\right)\right]=2\left(\hat{\boldsymbol t}_{ik}\cdot\hat{\boldsymbol r}_{ik}\right)s_{ik}$. The double-angle combination is invariant under the reversal $\hat{\boldsymbol t}_{ik}\rightarrow -\hat{\boldsymbol t}_{ik}$ and is parity odd, while changing sign under a spatial reflection. With the outward-tangent convention adopted here, $\chi_{ik}$ has the same handedness sign as $s_{ik}$ and differs only in its angular weighting. We therefore retain $s_{ik}$ for the present measurements and leave a direct comparison with $\chi_{ik}$ to future work.

We combine the radial segments of each arm into a single signed winding $h_i$, while its unsigned amplitude defines the arm spirality $S_i$:
\begin{equation}
h_i =
\frac{\sum_k w_{ik}s_{ik}}{\sum_k w_{ik}},
\qquad
S_i=|h_i|.
\label{eq:arm_spirality}
\end{equation}
After constructing each segment with the density weighting described in Sec.~\ref{sec:fans}, segments are combined using $w_{ik}=N_{{\rm ridge},ik}$, the number of ridge points contributing to segment $k$. By construction, $-1\leq h_i\leq1$ and $0\leq S_i\leq1$. Positive (negative) $h_i$ corresponds to coherent CCW (CW) winding. Under spatial reflection $h_i\rightarrow-h_i$, whereas $S_i$ is invariant. $S_i$ measures how strongly an arm departs from the radial direction while maintaining a common azimuthal winding sense across its segments: a radial arm has $S_i\simeq0$, whereas a coherently winding arm has increasingly larger $S_i$. Conversely, an arm may exhibit substantial local curvature or small-scale wiggles, yet have low spirality. This can occur when CW and CCW azimuthal contributions from successive segments offset one another.

At the population level, we separately retain the signed winding and its unsigned amplitude. We define
\begin{equation}
H_{\rm proj}
=
\frac{\sum_i W_i h_i}{\sum_i W_i},
\qquad
\mathcal{S}_{\rm proj}
=
\frac{\sum_i W_i |h_i|}{\sum_i W_i},
\label{eq:projected_estimators}
\end{equation}
where $W_i=N_{{\rm seg},i}^{\rm valid}$ gives larger weight to arms traced by more valid radial segments. $H_{\rm proj}$ measures the net signed winding, whereas $\mathcal{S}_{\rm proj}$ measures its unsigned amplitude. Under parity symmetry, $\langle H_{\rm proj}\rangle=0$, whereas $\mathcal{S}_{\rm proj}$ need not vanish because gravitational evolution can generate non-radial filament configurations without selecting a preferred handedness.

To test handedness independently of winding amplitude, we also use the sign-only projected asymmetry
\begin{equation}
A_{\rm sign}
=
\frac{N_{\rm CCW}-N_{\rm CW}}
     {N_{\rm CCW}+N_{\rm CW}}.
\label{eq:sign_asymmetry}
\end{equation}
The three statistics therefore retain complementary information: $H_{\rm proj}$ uses both winding sign and amplitude, $\mathcal{S}_{\rm proj}$ retains amplitude but discards sign, and $A_{\rm sign}$ retains sign but discards amplitude.

\textit{Extrinsic estimators.} — The observables above treat handedness as measured in the local sky plane of each arm. A purely projected average therefore tests whether CW and CCW winding are globally balanced across the observed sky. A coherent three-dimensional parity-odd pattern organized around a preferred cosmic axis, however, could appear with opposite projected signs when viewed from opposite sides of that axis \citep{2011PhLB..699..224L}. Such a signal could therefore cancel in $H_{\rm proj}$. Here, ``extrinsic'' means that the locally measured handedness is interpreted together with the viewing direction of its parent FANS relative to an external trial cosmic axis, allowing opposite projected signs to represent the same underlying three-dimensional winding pattern.

For an axial three-dimensional pattern, the projected handedness should vary approximately as a dipole across the sky. For each trial axis $\hat{\mathbf d}$, we therefore use $\mu_i(\hat{\mathbf d})=\hat{\mathbf n}_i\cdot\hat{\mathbf d}=\cos\theta_i$ as the dipolar angular coordinate, where $\theta_i$ is the angle between the line of sight $\hat{\mathbf n}_i$ to the parent FANS and the trial axis.

We test for this dipolar modulation by fitting the arm-level signed winding,
\begin{equation}
h_i =
H_0(\hat{\mathbf d})
+
A_{\rm dip}(\hat{\mathbf d})\,
\mu_i(\hat{\mathbf d})
+
\eta_i(\hat{\mathbf d}).
\label{eq:dipole_model}
\end{equation}
Here $h_i$ is the signed winding of arm $i$, $H_0$ is a fitted projected monopole, $A_{\rm dip}$ is the amplitude of the handedness dipole for the trial direction $\hat{\mathbf d}$, and $\eta_i$ denotes the residual arm-to-arm variation not described by the monopole-plus-dipole model. This residual includes intrinsic geometric scatter and reconstruction noise; it is not fitted as an additional arm-level parameter. Details of the weighted fit are given in Appendix~\ref{app:extrinsic}. Thus a positive $A_{\rm dip}$ corresponds, for the chosen representative of the axis, to preferentially positive handedness toward $+\hat{\mathbf d}$ and negative handedness toward $-\hat{\mathbf d}$.

We repeat this fit while scanning the trial axis $\hat{\mathbf d}$. For each direction, the FANS positions define a new set of $\mu_i(\hat{\mathbf d})$ and hence a fitted dipole amplitude $A_{\rm dip}(\hat{\mathbf d})$. The preferred axis $\hat{\mathbf d}_{\rm dip}$ is the direction that maximizes the magnitude of this dipolar modulation $|A_{\rm dip}(\hat{\mathbf d})|$. We maximize the absolute amplitude because $\hat{\mathbf d}$ and $-\hat{\mathbf d}$ describe the same physical axis and simply reverse the sign of $A_{\rm dip}$. We therefore scan only the positive-declination hemisphere. The redshift-bin measurements are evaluated at that catalogue's full-range best-fit axis.

As a complementary sign-only test, we fold the two viewing hemispheres about this same axis. Operationally, the projected handedness sign is reversed on one side of the axis so that opposite projected signs expected from the same underlying three-dimensional winding sense are counted together. We define $q_i(\hat{\mathbf d})\!=\!{\rm sgn}\!\left[h_i\,\mu_i(\hat{\mathbf d})\right]$ and the axial projected asymmetry
\begin{equation}
A_{\rm ax}(\hat{\mathbf d})
\!=\!
\frac{
N^{\rm ax}_{\rm CCW}(\hat{\mathbf d})
\!-\!
N^{\rm ax}_{\rm CW}(\hat{\mathbf d})
}{
N^{\rm ax}_{\rm CCW}(\hat{\mathbf d})
\!+\!
N^{\rm ax}_{\rm CW}(\hat{\mathbf d})
}.
\label{eq:axial_sign}
\end{equation}
Here $N^{\rm ax}_{\rm CCW}$ and $N^{\rm ax}_{\rm CW}$ count arms with $q_i=+1$ and $q_i=-1$, respectively; CW and CCW therefore refer to the winding sense after folding the two viewing hemispheres relative to the chosen representative of the axis. The sign convention follows the positive-declination representative adopted for $\hat{\mathbf d}_{\rm dip}$. We evaluate $A_{\rm ax}$ only at the axis selected by the continuous-handedness dipole, $A_{\rm ax}\equiv A_{\rm ax}(\hat{\mathbf d}_{\rm dip})$, rather than performing a second directional optimization. Thus $A_{\rm dip}$ uses the continuous signed winding to determine the axis, whereas $A_{\rm ax}$ provides a sign-only test of the same axial organization.

\section{Results}
\label{sec:results}
\vspace{0.2cm}

\begin{figure*}[!ht]
\centering
\includegraphics[width=1\linewidth]{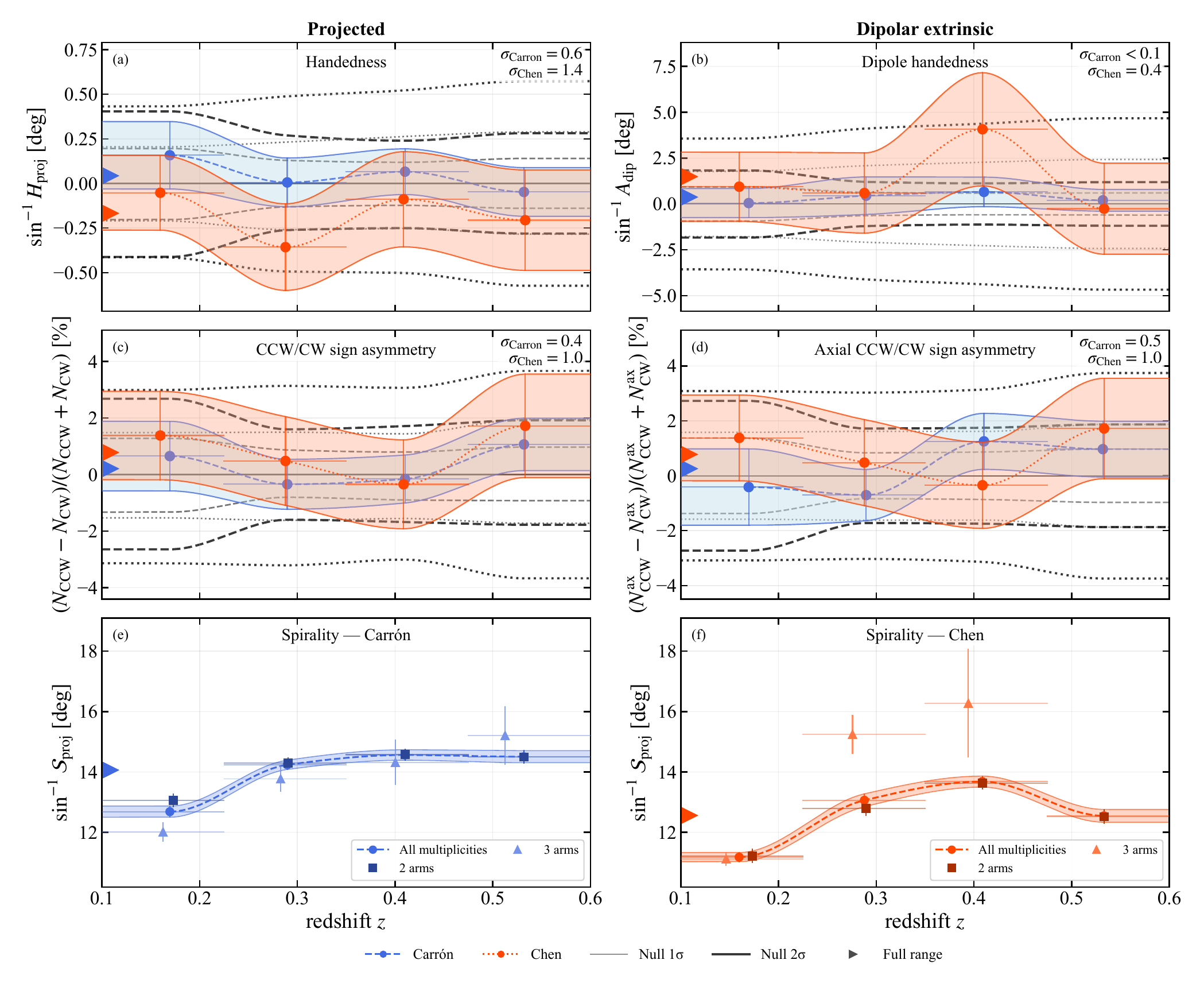}
\caption{The upper two rows show projected handedness $H_{\rm proj}$ and sign asymmetry (left), and the extrinsic handedness dipole $A_{\rm dip}$ and axial sign asymmetry (right). The bottom row shows spirality for all accepted FANS and the two- and three-arm subsamples, separately for the Carr\'on sample (left) and the Chen sample (right). Blue dashed and red dotted curves denote the Carr\'on and Chen samples, respectively. Markers are centred at the mean redshift of each bin, horizontal bars span the bin boundaries, and vertical bars show the $1\sigma$ uncertainties obtained from a spatial block bootstrap with $N_{\rm side}=16$. Triangles at the left margin show the measurements over the full unbinned range $0.1<z<0.6$. In the parity-sensitive panels, grey envelopes indicate the $1\sigma$- and $2\sigma$-equivalent ranges of the block sign-flip null ensembles, with dashed boundaries for the Carr\'on sample and dotted boundaries for the Chen sample. The quoted significances refer to the full-range unbinned measurements. The continuous observables are expressed in degrees by applying $\arcsin(x)$. The final Chen three-arm point is omitted because too few objects are available for a reliable estimate. The fitted axes are $(\alpha,\delta)=(329^\circ,62^\circ)$ for the Carr\'on sample and $(198^\circ,50^\circ)$ for the Chen sample; neither dipole is statistically significant, with global significances of $<0.1\sigma$ and $0.4\sigma$ for the full-range handedness dipole amplitude $|A_{\rm dip}|$ in the Carr\'on and Chen samples, respectively.}
\label{fig:results}
\end{figure*}

Using the fiducial FANS construction parameters, we measure the projected and axis-conditioned statistics. The results are summarized in Fig.~\ref{fig:results}: the upper two rows compare projected measurements on the left with dipolar extrinsic measurements on the right, while the bottom row shows spirality for the full and fixed-multiplicity samples. The continuous quantities $H_{\rm proj}$, $\mathcal{S}_{\rm proj}$, and $A_{\rm dip}$ are  expressed in degrees by applying $\arcsin(x)$.

Uncertainties are estimated from 1000 bootstrap resamples of HEALPix blocks with $N_{\rm side}=16$, keeping neighbouring arms within each selected block together and sharing resamples across bins and multiplicity classes within each catalogue. Significances are calibrated using 1000 block sign-flip nulls, in which handedness signs are flipped coherently within spatial blocks while preserving the observed positions, redshifts, weights, spatial grouping, and winding amplitudes. Every null realization is passed through the same inference procedure as the data, including the complete directional scan for the extrinsic estimators. Each extrinsic bootstrap and null realization reselects one full-range axis, used for all its bins. This empirical procedure accounts for within-block correlations, survey geometry, and the directional look-elsewhere effect without requiring a model for the absolute FANS morphology.

Both projected handedness estimators are consistent with parity symmetry. The binned values of $H_{\rm proj}$ fluctuate around zero without a persistent preference for CW or CCW winding, and the full-range deviations from the sign-flip null are   $0.6\sigma$ for the Carrón sample and $1.4\sigma$ for the Chen sample. The corresponding significances for $A_{\rm sign}$ are   $0.4\sigma$ and $1.0\sigma$, respectively. The same qualitative null behaviour persists across the 27 construction choices examined in Appendix~\ref{app:robustness}.

The extrinsic measurements likewise show no significant handedness organization. Full-range $A_{\rm dip}$ significances are below $0.1\sigma$ for the Carrón sample and $0.4\sigma$ for the Chen sample; $A_{\rm ax}$ gives $0.5\sigma$ and $1.0\sigma$, respectively. The selected directions are reported in Fig.~\ref{fig:results}; neither dipole is statistically significant, and no preferred axis is detected.

In contrast, $\mathcal{S}_{\rm proj}$ shows a net increase from the lowest- to the highest-redshift bin in both fiducial samples, indicating stronger coherent azimuthal winding at earlier cosmic times. We quantify this evolution through the angular spirality contrast $\Delta\mathcal{S}=(180^\circ/\pi)[\arcsin\mathcal{S}_{\rm proj}(z_4)-\arcsin\mathcal{S}_{\rm proj}(z_1)]$, where $z_1$ and $z_4$ denote the first and last redshift bins. The trend persists for two-arm FANS: $\Delta\mathcal{S}^{(2)}=(1.44\pm0.27)^\circ$ for the Carrón sample and $(1.30\pm0.28)^\circ$ for the Chen sample, with 95\% percentile intervals $[0.92^\circ,1.96^\circ]$ and $[0.74^\circ,1.81^\circ]$, respectively (Appendix~\ref{app:multiplicity}). The Carrón three-arm sample also shows a positive contrast, $\Delta\mathcal{S}^{(3)}=(3.20\pm1.00)^\circ$, with a 95\% interval $[1.28^\circ,5.03^\circ]$; the Chen three-arm sample lacks adequate support in the last bin. These results support spirality evolution within different multiplicity classes, rather than a trend produced solely by their changing mixture. The evolution is not universally monotonic: the Chen sample peaks in the third bin before declining. Nevertheless, positive contrasts prevail across the construction grid (26/27 for the full Carrón sample and 25/27 for the full Chen sample), and none of the supported negative contrasts is statistically significant at the 95\% level (Appendix~\ref{app:robustness}).

\section{Discussion and conclusions}
\label{sec:conclusion}
\vspace{0.2cm}

This Letter establishes Filament-Arm Node Systems (FANS) as a new class of cosmic-web objects and presents the first measurements of their winding geometry through spirality and handedness. Across two distinct SDSS filament reconstructions, all projected and extrinsic handedness estimators are consistent with parity symmetry.  No persistent CW or CCW preference emerges across the 27 variations of construction parameters examined here.

FANS recast connected filament ridges as node-centred objects with well-defined radial and azimuthal directions, allowing the same structures to encode both a parity-odd signed winding and a parity-even spirality amplitude. This dual description complements galaxy-spin studies, which commonly compress apparent spiral morphology into a binary winding label, and higher-order density statistics, which do not condition the measurement on individual node environments. Beyond parity tests, spirality provides a continuous large-scale morphological indicator that can be studied as a function of redshift, node mass, connectivity, environment, filament reconstruction, and cosmological model.

The increase of $\mathcal{S}_{\rm proj}$ between the lowest- and highest-redshift bins over $0.1<z<0.6$, quantified by the spirality contrast $\Delta\mathcal{S}$ and also found within individual arm-multiplicity classes, suggests that FANS arms are more strongly wound at earlier times and tend to become straighter and more radially oriented as the cosmic web evolves. This interpretation is qualitatively supported by simulations showing late-time filament straightening and, for sufficiently long filaments, longitudinal expansion \citep{Ilc:2024rrm,Galarraga-Espinosa:2023zmv}. Targeted $N$-body and hydrodynamical simulations can now test whether this inferred evolution follows from standard gravitational growth and calibrate its dependence on node mass, connectivity, and tidal environment.

The observational reach of FANS should expand rapidly with the next generation of wide-field surveys. DESI alone provides a spectroscopic tracer sample roughly an order of magnitude larger than previous SDSS programmes, while Euclid, Rubin, CSST, and PFS will add substantially denser and deeper samples \citep{DESI:2023dwi,LSST:2008ijt,Euclid:2024yrr,CSST:2025ssq,PFSTeam:2012fqu}. Combining the larger tracer samples, we expect at least an order-of-magnitude increase in the number of usable systems as a conservative expectation, while the increased depth will extend the measurement to earlier cosmic epochs. At characteristic scales of $\sim60\,h^{-1}{\rm Mpc}$, FANS provide a direct bridge between cosmic-web winding and fundamental symmetry tests. The measurements presented here establish the first benchmark for this framework and lay the foundation for a new precision probe of cosmic-web symmetry, morphology, and gravitational evolution.

\begin{acknowledgments}
The authors would like to acknowledge the support from the start-up funding of Zhejiang University and Zhejiang provincial top level research support program, and the use of the SilkRiver Supercomputer of Zhejiang University (China). This work made use of the catalogues of \citet{Duque:2021xgw,Chen:2015oqa}.
\end{acknowledgments}\vspace{-0.1cm}

\vspace{0.1cm}
\begin{contribution}
PDSF conceived the study; developed the methodology; performed the statistical analysis; contributed to the interpretation; drafted the manuscript; and assembled the data products. RC advised on the statistical analysis and physical interpretation and critically revised the manuscript.
\end{contribution}

\section*{Data and code availability}

The code used to construct the FANS catalogs and perform the analysis presented in this work is publicly available at \url{https://github.com/pdsferreira/fans}.
\appendix
\renewcommand{\theHequation}{\Alph{section}.\arabic{equation}}

\section{Additional details of the FANS estimators}
\label{app:extrinsic}

\subsection{Weighted dipole fit and monopole leakage}

For each trial direction $\hat{\mathbf d}$, we fit the dipole model in Eq.~(\ref{eq:dipole_model}) by minimizing the weighted residual sum $\sum_i W_i\eta_i^2(\hat{\mathbf d})$, using the same arm weights $W_i=N_{{\rm seg},i}^{\rm valid}$ adopted for the projected estimators. To separate the fitted dipole from any projected monopole, we define the weighted means
\begin{equation}
\bar h_W
=
\frac{\sum_i W_i h_i}{\sum_i W_i},
\qquad
\bar\mu_W(\hat{\mathbf d})
=
\frac{
\sum_i W_i\mu_i(\hat{\mathbf d})
}{
\sum_i W_i
}.
\label{eq:weighted_means}
\end{equation}
With the adopted weights, $\bar h_W=H_{\rm proj}$. For compactness, the dependence of $\mu_i$ and $\bar\mu_W$ on $\hat{\mathbf d}$ is suppressed in the equations below. The weighted least-squares solution is
\begin{equation}
\begin{aligned}
A_{\rm dip}(\hat{\mathbf d})
&=
\frac{
\sum_i W_i
\left(\mu_i-\bar\mu_W\right)
\left(h_i-\bar h_W\right)
}{
\sum_i W_i
\left(\mu_i-\bar\mu_W\right)^2
},\\
H_0(\hat{\mathbf d})
&=
\bar h_W
-
A_{\rm dip}(\hat{\mathbf d})\bar\mu_W .
\end{aligned}
\label{eq:weighted_dipole_solution}
\end{equation}

The fitted residuals are $\eta_i(\hat{\mathbf d})=h_i-H_0(\hat{\mathbf d})-A_{\rm dip}(\hat{\mathbf d})\mu_i(\hat{\mathbf d})$. No parametric distribution is assumed for these residuals, since the statistical calibration of the fitted dipole is obtained empirically from the sign-flip null realizations. The intercept is important because an incomplete and anisotropic survey footprint need not sample the two sides of a trial axis symmetrically. If the fit were forced through the origin, the estimator would instead be $\widetilde A_{\rm dip}=\sum_i W_i\mu_i h_i \big/\sum_i W_i\mu_i^2$. For the model in Eq.~(\ref{eq:dipole_model}), and assuming that the residuals have no systematic projection onto the trial dipole, $\left\langle\sum_i W_i\mu_i\eta_i\right\rangle = 0$, the expectation of the no-intercept estimator is
\begin{equation}
\left\langle
\widetilde A_{\rm dip}
\right\rangle
=
A_{\rm dip}
+
H_0
\frac{
\sum_i W_i\mu_i
}{
\sum_i W_i\mu_i^2
}.
\label{eq:monopole_leakage}
\end{equation}
Thus a non-zero weighted mean of $\mu_i$, as can arise from an incomplete or anisotropic footprint and finite sampling, allows monopole-to-dipole leakage unless the intercept is included.

\subsection{Axis scan and trial-factor calibration}

Because reversing a trial direction exchanges the two viewing hemispheres, $\mu_i\rightarrow-\mu_i$ and $A_{\rm dip}\rightarrow-A_{\rm dip}$. The pair $\{\hat{\mathbf d},-\hat{\mathbf d}\}$ therefore represents a single physical axis, so we scan only one hemisphere and use $|A_{\rm dip}|$ as the axis-selection statistic.

Selecting the direction that maximizes the observed dipole introduces a directional look-elsewhere effect. We account for this by repeating the complete axis search in every null realization $r$ and recording the trial maximum dipole $T_r^{\rm dip}=\max_{\hat{\mathbf d}}|A_{{\rm dip},r}(\hat{\mathbf d})|$. The observed statistic is defined analogously as $T_{\rm obs}^{\rm dip}=\max_{\hat{\mathbf d}}|A_{\rm dip}^{\rm obs}(\hat{\mathbf d})|$. The empirical global $p$-value is
\begin{equation}
p_{\rm global}^{\rm dip}
=
\frac{
1+
\sum_{r=1}^{N_{\rm null}}
\mathbf{1}\!\left[
T_r^{\rm dip}
\geq
T_{\rm obs}^{\rm dip}
\right]
}{
N_{\rm null}+1
},
\label{eq:global_pvalue}
\end{equation}
where $\mathbf{1}[\cdot]$ is the indicator function. Repeating the full directional optimization in every null realization automatically propagates the survey footprint, the discrete axis scan, and the associated look-elsewhere effect.

The axial sign statistic does not introduce a second axis search. In each null realization, we first determine the axis selected by its continuous-handedness dipole, $\hat{\mathbf d}_{{\rm dip},r}=\arg\max_{\hat{\mathbf d}}\,|A_{{\rm dip},r}(\hat{\mathbf d})|$, and then evaluate $A_{{\rm ax},r}=A_{{\rm ax},r}(\hat{\mathbf d}_{{\rm dip},r})$. The same positive-declination representative convention adopted for the data is applied to every null realization. The empirical null distribution of $A_{\rm ax}$ therefore propagates the same axis-selection procedure used for the data without introducing an additional directional trial factor.

The directional search uses HEALPix pixel centres with $N_{\rm side}=32$ over the adopted hemisphere. For $H_{\rm proj}$, $A_{\rm sign}$, and $A_{\rm ax}$, empirical two-sided $p$-values are reported as Gaussian-equivalent significances $Z=\Phi^{-1}(1-p/2)$, where $\Phi$ is the standard-normal cumulative distribution. The optimized dipole statistic $T_{\rm obs}^{\rm dip}=\max_{\hat{\mathbf d}}|A_{\rm dip}(\hat{\mathbf d})|$ is non-negative and tested in its upper tail, giving $Z=\Phi^{-1}(1-p_{\rm global}^{\rm dip})$.

\subsection{Multiplicity selection and redshift contrasts}
\label{app:multiplicity}

The diagnostics in Fig.~\ref{fig:fans_redshift} compare arm multiplicities before and after the tracing quality cuts, with the survey-boundary cut held fixed. These transitions include nodes subsequently rejected for having fewer than two surviving arms.

For arms $\mathcal{A}_{m,b}$ belonging to accepted FANS of observed multiplicity $m$ in redshift bin $b$, we measure

\begin{equation}
\begin{aligned}
\!\!\mathcal{S}_{\rm proj}^{(m)}(z_b)&\!=\!\frac{\sum_{i\in\mathcal{A}_{m,b}}W_i|h_i|}{\sum_{i\in\mathcal{A}_{m,b}}W_i},\\
\!\!\Delta\mathcal{S}^{(m)}&\!=\!\frac{180^\circ}{\pi}\!\Bigl[\!\arcsin\mathcal{S}_{\rm proj}^{(m)}(\!z_4\!)\!-\!\arcsin\mathcal{S}_{\rm proj}^{(m)}(\!z_1\!)\!\Bigr].
\end{aligned}
\label{eq:multiplicity_contrast}
\end{equation}

The redshift-bin edges are those of Sec.~\ref{sec:fans}. The bootstrap uses the same block multiplicities across bins and classes. In each realization, we apply $\arcsin$ separately to the two endpoint values before subtracting and converting to degrees; the standard deviation and percentiles of these angular contrasts give the error and interval, including bin covariance. We require at least ten FANS in five occupied blocks per bin for supported conditional comparisons; finite estimates below this threshold are diagnostic only. Sign flips leave $|h_i|$ unchanged and are not used to test the redshift dependence of spirality.

\section{Robustness to FANS construction parameters}
\label{app:robustness}

\begin{figure*}[!ht]
\centering
\includegraphics[width=0.87\linewidth]{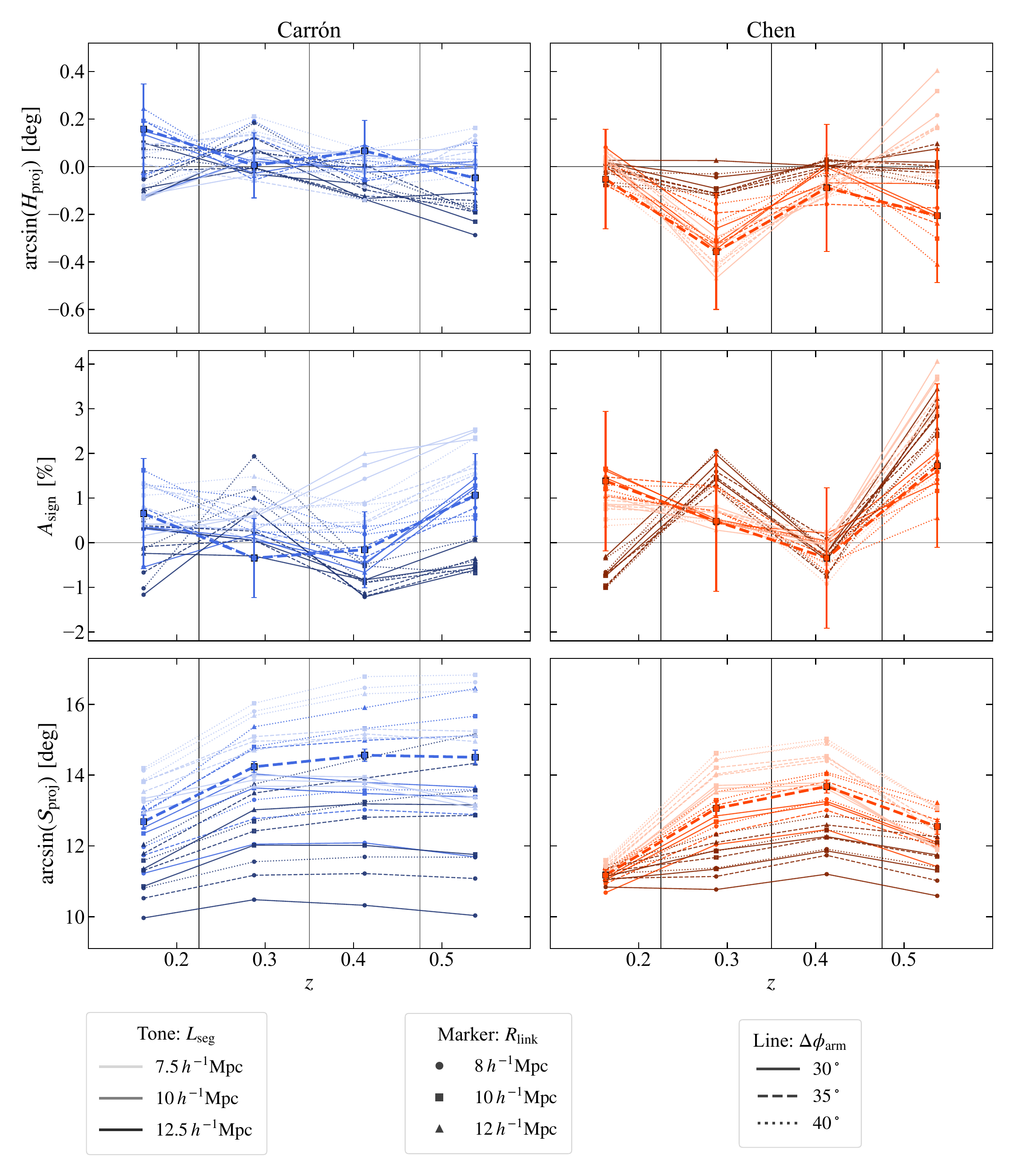}
\caption{Projected handedness $H_{\rm proj}$, sign asymmetry $A_{\rm sign}$, and spirality $\mathcal{S}_{\rm proj}$ for the 27 combinations of $R_{\rm link}$, $\Delta\phi_{\rm arm}$, and $L_{\rm seg}$ considered in the robustness analysis. Error bars are shown only for the fiducial construction.  Vertical lines mark the boundaries of the four redshift bins.}
\label{fig:robustness}
\end{figure*}

\begin{figure*}[!ht]
\centering
\includegraphics[width=0.95\linewidth]{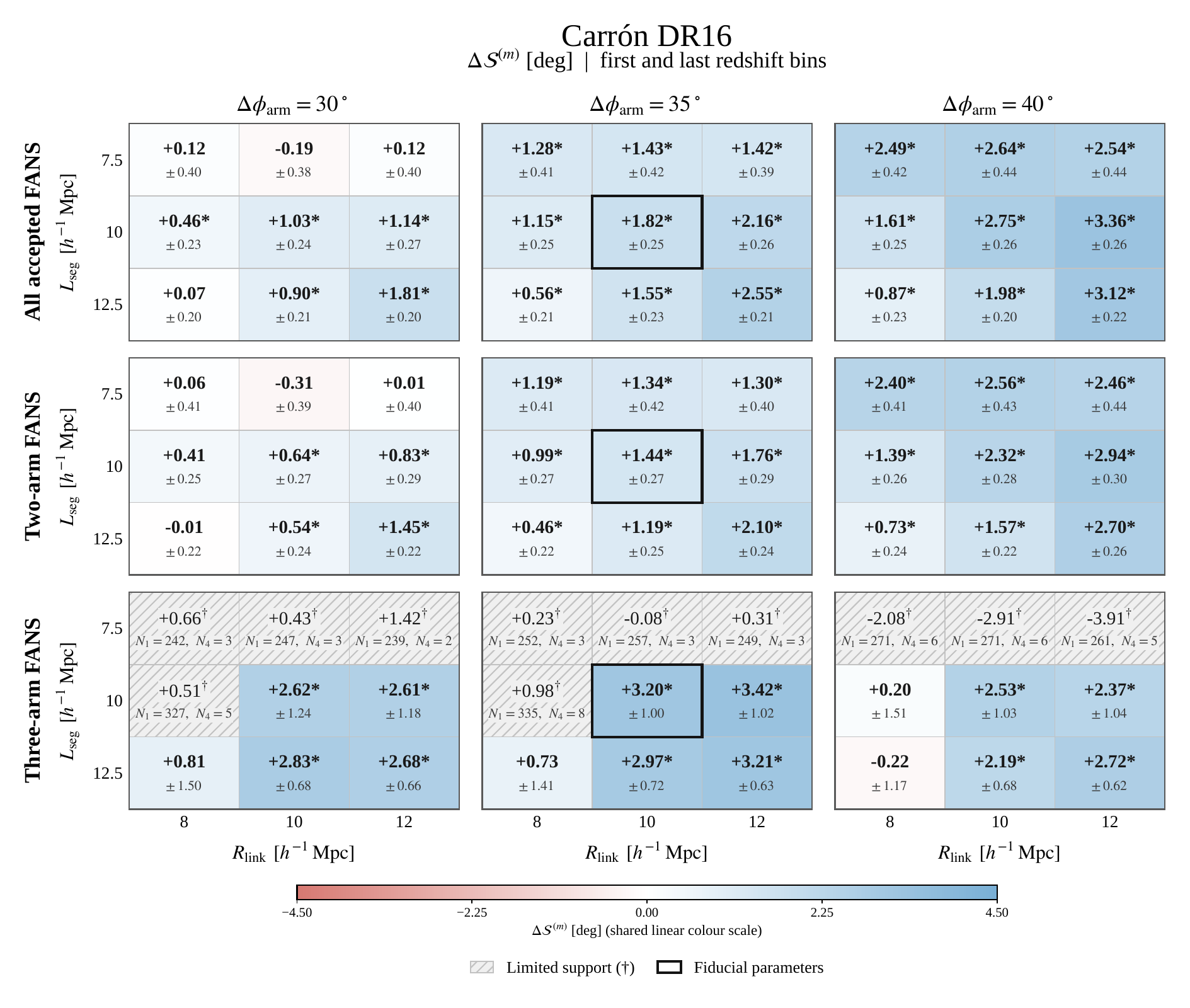}
\caption{Spirality contrast $\Delta\mathcal{S}=(180^\circ/\pi)[\arcsin\mathcal{S}_{\rm proj}(z_4)-\arcsin\mathcal{S}_{\rm proj}(z_1)]$ between the last and first redshift bins for the Carr\'on sample across the 27 constructions and three multiplicity selections. Cells give contrasts in degrees and paired-bootstrap standard errors; asterisks mark 95\% percentile intervals excluding zero. Colour intensity encodes the signed contrast on a common scale with Fig.~\ref{fig:delta_s_chen}. Hatched cells have insufficient support or an undefined interval, with endpoint counts $N_1,N_4$ shown; dashes denote undefined contrasts. Black outlines mark the fiducial construction.}
\label{fig:delta_s_carron}
\end{figure*}

\begin{figure*}[!ht]
\centering
\includegraphics[width=0.95\linewidth]{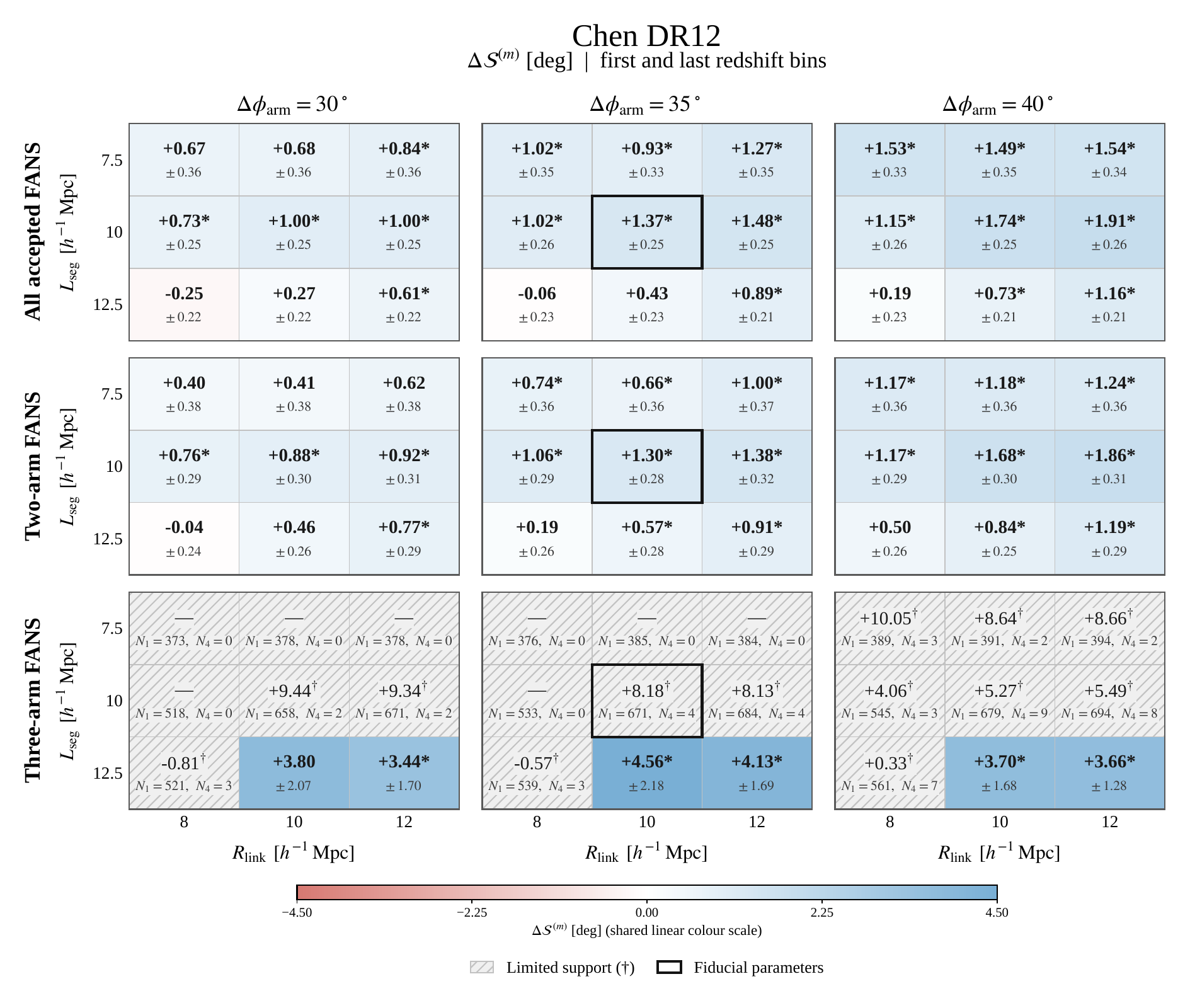}
\caption{Same as Fig.~\ref{fig:delta_s_carron}, for the Chen sample.}
\label{fig:delta_s_chen}
\end{figure*}

We test the sensitivity of FANS measurements to arm-construction choices by reconstructing FANS and recomputing their projected statistics for 27 parameter combinations: $R_{\rm link}\in\{8,10,12\}\,h^{-1}{\rm Mpc}$, $\Delta\phi_{\rm arm}\in\{30^\circ,35^\circ,40^\circ\}$, and $L_{\rm seg}\in\{7.5,10,12.5\}\,h^{-1}{\rm Mpc}$. The fiducial construction uses $R_{\rm link}=10\,h^{-1}{\rm Mpc}$, $\Delta\phi_{\rm arm}=35^\circ$, and $L_{\rm seg}=10\,h^{-1}{\rm Mpc}$. Fig.~\ref{fig:robustness} compares $H_{\rm proj}$, $A_{\rm sign}$, and $\mathcal{S}_{\rm proj}$ across the grid for both filament reconstructions. To facilitate comparison, only fiducial error bars are shown, estimated using a HEALPix block bootstrap with $N_{\rm side}=16$.  The number of radial segments remains three, so varying $L_{\rm seg}$ changes the maximum tracing radius to $R_{\rm arm}=3L_{\rm seg}$. The seed-grouping threshold remains fixed. Narrower tracing angles or shorter segments increase sensitivity to fragmentation and sampling, whereas broader angles or longer segments can mix neighbouring ridge structures or smooth their curvature. We repeat the multiplicity-conditioned projected measurements for each construction.

We restrict this grid test to the projected estimators because it targets the arm-construction stage shared by the projected and extrinsic analyses. The extrinsic estimators require no separate reconstruction: $A_{\rm dip}$ uses the same arm-level windings $h_i$ and weights $W_i$ as its projected counterpart (Eq.~\ref{eq:weighted_dipole_solution}), while $A_{\rm ax}$ uses the same handedness signs after hemispheric folding (Eq.~\ref{eq:axial_sign}). Their additional angular dependence enters through $\mu_i(\hat{\mathbf d})=\hat{\mathbf n}_i\cdot\hat{\mathbf d}$, determined by the parent node position and the trial axis. The projected statistics therefore provide complementary diagnostics of construction-induced changes in the signed mean, sign balance, and winding amplitude underlying both analyses. Repeating the extrinsic inference across the grid requires a directional search in every sign-flip null realization (Appendix~\ref{app:extrinsic}), making it substantially more computationally expensive than the projected analysis in our implementation. We thus assess shared construction-level sensitivity while retaining the fiducial construction for optimized extrinsic measurements and their statistics.

The parity-sensitive projected observables remain qualitatively stable across the construction grid, with no persistent CW or CCW preference. Projected spirality $\mathcal{S}_{\rm proj}$ shows a broadly consistent increase from the lowest- to the highest-redshift bin, although its normalization and curve shape depend on the construction parameters. The full-sample contrast is positive in 26/27 constructions for the Carrón sample and 25/27 for the Chen sample, with 95\% intervals entirely above zero in 23/27 and 20/27, respectively. For two-arm FANS, the corresponding counts are 25/27 and 26/27 positive contrasts, of which 22/27 and 20/27 have intervals entirely above zero. Every negative contrast in these comparisons is statistically consistent with zero at the 95\% level; none provides significant evidence for the opposite evolutionary trend. Figures~\ref{fig:delta_s_carron} and \ref{fig:delta_s_chen} give the contrasts and support diagnostics for all, two-arm, and three-arm selections.

For the Chen sample, both the full and two-arm selections show an increase up to the third bin followed by a decline in the highest-redshift bin in all 27 constructions. The peak position in the full Carrón sample depends on construction, so the increase is not universally monotonic. Reducing $R_{\rm link}$ generally weakens the redshift dependence, particularly for the Chen sample, indicating sensitivity to how closely the initial ridge--node connection must approach the node. Robustness therefore concerns the predominance of a positive redshift contrast, not a parameter-independent normalization or curve shape. Because $\mathcal{S}_{\rm proj}$ measures coherent azimuthal departure from the radial direction, this behaviour suggests stronger coherent winding at earlier cosmic times. Such evolution may be associated with gravitational straightening, and possibly longitudinal stretching, of filament arms, making them more radial toward the present.

\FloatBarrier 
\bibliography{references}{}
\bibliographystyle{aasjournalv7}

\end{document}